\documentclass[twocolumn,showpacs,preprintnumbers,amsmath,amssymb,nofootinbib]{revtex4}
\usepackage{graphicx}% Include figure files
\usepackage{dcolumn}% Align table columns on decimal point
\usepackage{bm}% bold math
\usepackage{color}

\newcommand{\eq}[1]{Eq.~(\ref{#1})}
 
\newcommand{\bb}{\ensuremath{B\!-\!\Bbar{}\,}}

\newcommand{\bbm}{\bb\ mixing}

\newcommand{\Bbar}{\,\overline{\!B}}

\begin{document}

\title{$B\to X_d \gamma$ and constraints on new physics}
% Force line breaks with \\
%
\author{Andreas Crivellin and Lorenzo Mercolli}
\affiliation{Albert Einstein Center for Fundamental Physics, Institute for Theoretical Physics,\\
              University of Bern, CH-3012 Bern, Switzerland.}
             % \date{\today}
%
\begin{abstract}
We combine recent progress in measuring the branching ratio of the decay $B\to X_d \gamma$ with the discovery that hadronic uncertainties in the CP-averaged branching ratio drop out to a large extent. Implications of these improvements on the size of possible new physics effects are investigated. We find the updated SM prediction for the CP-averaged branching ratio to be $\langle\mathrm{Br}\left[ B\to X_d\gamma \right]^\mathrm{SM}_{E_{\gamma}>1.6 {\rm GeV}}\rangle = \;1.54^{+0.26}_{-0.31}\times10^{-5}$, which should be compared with the experimental value of $\langle\mathrm{Br}\left[ B \to X_d \gamma \right]^\mathrm{exp}_{E_{\gamma}>1.6 {\rm GeV}}\rangle=(1.41\pm0.57) \times 10^{-5}$. 
After performing a model independent analysis, we consider different new physics models: the MSSM with generic sources of flavor violation, the two Higgs doublet model of type III and a model with right-handed charged currents. It is found that the constraints on the SUSY parameters $\delta^{d}_{13}$ have improved and that the absolute value of the right-handed quark mixing matrix element $\left|V^R_{td}\right|$ must be smaller than $1.5\times10^{-4}$. 
\end{abstract}

\maketitle

\section{\label{sec:level1}Introduction}

In the past, the main focus has been on the inclusive decay $B\to X_s \gamma$ while its analog with a down quark in the final state, $B\to X_d \gamma$, received much less attention. The reason for this was that both the experimental measurement ${\rm{Br}}[B\to X_s \gamma]^{{\rm exp}}_{E_{\gamma}>1.6 {\rm GeV}}=(3.60\pm0.23)\times10^{-4}$ \cite{Nakamura:2010zzi} and the standard model prediction (it is now known to NNLO precision) ${\rm{Br}}[B\to X_s \gamma]^{{\rm SM}}_{E_{\gamma}>1.6 {\rm GeV}}=(3.15\pm0.23)\times10^{-4}$ \cite{Misiak:2006zs,Misiak:2006ab} of this decay were significantly better compared to $B\to X_d \gamma$. However, this situation has changed recently:

\begin{list}{\labelitemi}{\leftmargin=1em}

\item The new CP-averaged branching ratio $ \langle\mathrm{Br}\left[ B \to X_d \gamma \right]^\mathrm{exp}_{E_{\gamma}>1.6 {\rm GeV}}\rangle=(1.41\pm0.57) \times 10^{-5}$ of the BABAR collaboration \cite{BABAR:2010ps,Wang:2011sn} (CP averaging is denoted by $\langle...\rangle$ throughout this article) is more precise than the previous one and the photon cut is lower which reduces the error of the extrapolation to 1.6 GeV\footnote{Note that in the ICHEP 2010 update of HFAG the value for $B\to X_d \gamma$ is not extrapolated from the photon cut of 2.26 GeV used in the BABAR measurement to a cut of 1.6 GeV even though this was done in previous updates. It can thus be misleading to compare the value of HFAG with the one of PDG (quoted in the HFAG analysis) since the latter one has been extrapolated down to 1.6 GeV. We obtained the value quoted above by using the extrapolation of HFAG. In order to be conservative we doubled the error given in {\itshape www.slac.stanford.edu/xorg/hfag/rare/ichep10/radll/btosg.pdf}.}. Furthermore, there are good experimental prospects for this decay: the analysis of existing BELLE data and the future super-B factories \cite{O'Leary:2010af,Abe:2010sj} will allow for a more precise determination of this branching ratio.

\item The theory prediction for the standard model (SM) contribution has been calculated in Ref.~\cite{Ali:1992qs} and the NLO QCD corrections can be found in Ref.~\cite{Ali:1998rr}. As in the case of $B\to X_s\gamma$, also $B\to X_d\gamma$ suffers from hadronic uncertainties but for the latter the non-perturbative contributions from up-quark loops are not CKM-suppressed which magnifies the error of the theory prediction. However, it has been only recently realized that most of these uncertainties drop out in the CP-averaged branching ratio \cite{Benzke:2010js,Hurth:2010tk}. Thus, the SM prediction for $B\to X_d\gamma$ can in principle be calculated with the same accuracy as $B\to X_s\gamma$.

\item In addition, the error in the determination of the CKM element $V_{td}$ has constantly decreased in the last years \cite{Charles:2004jd,Ciuchini:2000de}. This further reduces the uncertainty of the SM contribution to $B\to X_d\gamma$ which depends quadratically on $V_{td}$. The uncertainty coming from the determination of $V_{td}$ now only induces an error in the SM branching ratio of approximately $10 \%$ if one varies the value of $V_{td}$ within its $95\%{\rm CL}$ region.

\end{list}
\medskip

These significant improvements and promising prospects on the theoretical as well as on the experimental side motivate us to perform an updated analysis of $B\to X_d\gamma$ and the constraints placed from this decay. 

\begin{figure*}
\centering
\includegraphics[width=0.39\textwidth]{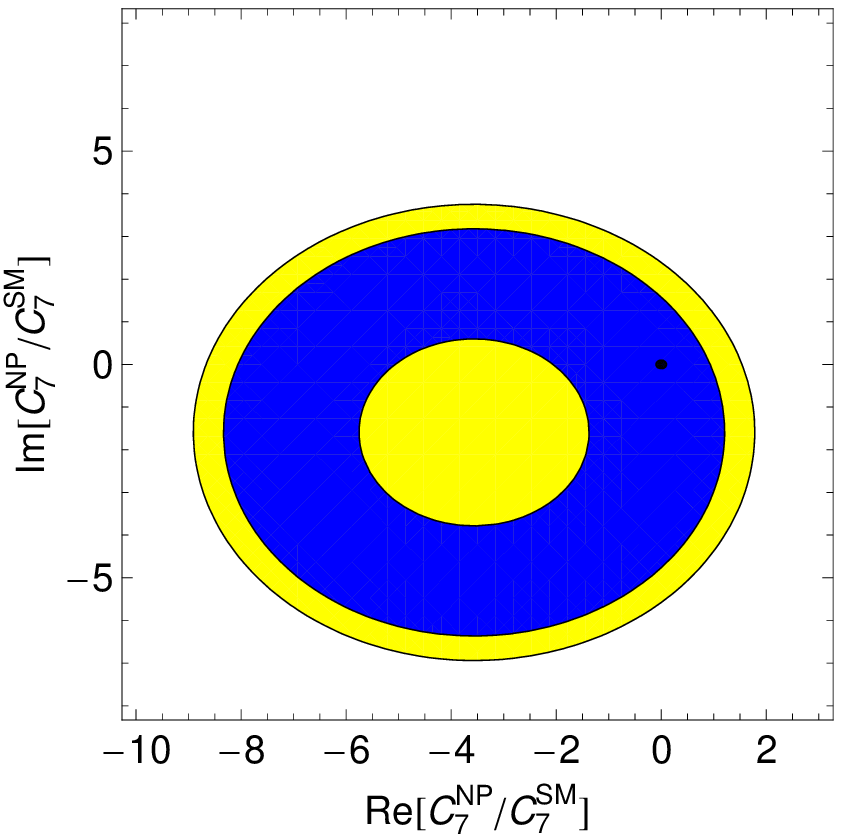}
\hspace{1cm}
\includegraphics[width=0.4\textwidth]{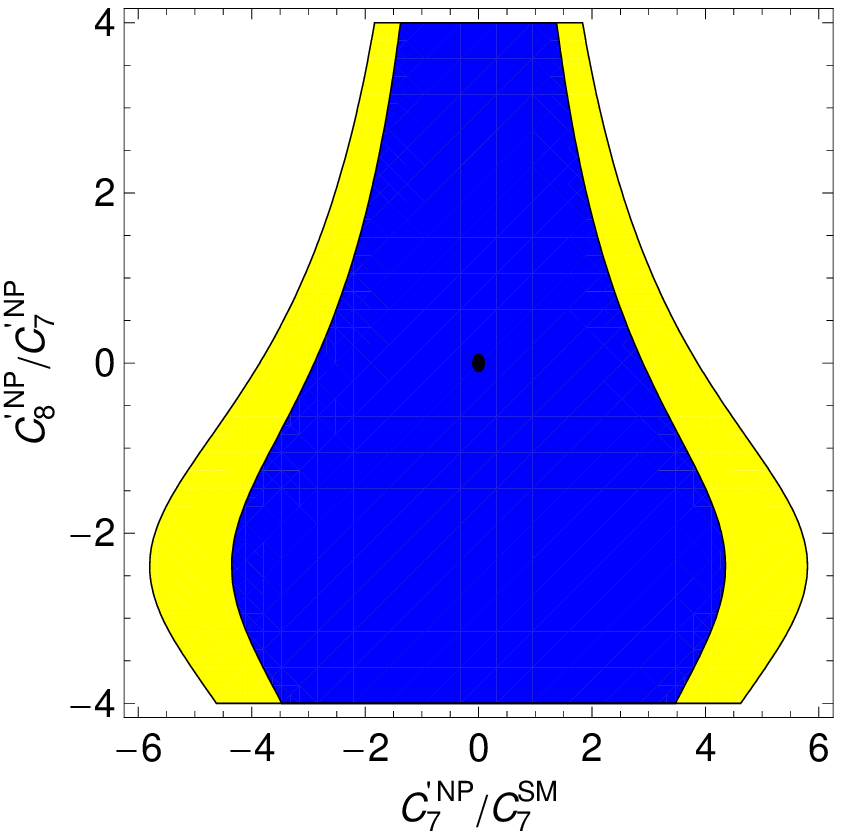}
\caption{The blue (yellow) region agrees with the measured branching ratio at the $1\sigma$ ($2\sigma$) level.
Left plot: Allowed region in the $Re\left[C^{\mathrm{NP}}_{7}/C^{\mathrm{SM}}_{7}\right]$--$Im\left[C^{\mathrm{NP}}_{7}/C^{\mathrm{SM}}_{7}\right]$ plane for $C^{\mathrm{NP}}_7/{C^{\mathrm{NP}}_8}=C^{\mathrm{SM}}_7/{C^{\mathrm{SM}}_8}$ and $C^{\prime\mathrm{NP}}_{7,8}=0$. 
Right plot: Allowed region in the $C^{\prime\mathrm{NP}}_{7}/C^{\mathrm{SM}}_{7}$--$C^{\prime\mathrm{NP}}_{8}/C^{\prime\mathrm{NP}}_{7}$ plane for $C^{\prime\mathrm{NP}}_{8}/C^{\prime\mathrm{NP}}_{7} \in \mathbb{R} $. Note that the constraints in the right plot are independent of the phase of $C^{\prime\mathrm{NP}}_{7}$. In both plots the SM point is marked by a black dot.
\label{C7}}
\end{figure*}

\section{Effective Hamiltonian and SM prediction}

In the SM the effective Hamiltonian governing $\bar{B}\to X_d\gamma$ is given by:

\begin{equation}
	\mathcal{H}_{\mathrm{eff}}=\dfrac{-4G_F}{\sqrt{2}}V_{td}^\star V_{tb} \left[ \sum_{i=1}^8 C_i O_i  + \epsilon_d \sum_{i=1}^2 C_i\left(O_i  - O_i^{u}\right) \right]\;,
	\label{Heff}
\end{equation}
where $O_1^u$, $O_2^u$, $O_1$, ...., $O_6$ are four-quark operators, $\epsilon_d=\frac{V_{ud}^*V_{ub}}{V_{td}^*V_{tb}}$ and the (chromo)magnetic operator ($O_{8}$) $O_{7}$ is given by
\begin{equation}
\begin{split}
	O_7 &= \dfrac{e}{16\pi^2}m_b(\mu)(\overline{d}_L\sigma_{\mu\nu}b_R)F^{\mu\nu}  \;, \\
  O_8 &= \dfrac{g_s}{16\pi^2}m_b(\mu)(\overline{d}_LT^a\sigma_{\mu\nu}b_R)G^{a\mu\nu} \;.
\end{split}
\end{equation}
In the presence of new physics (NP), additional operators may appear. We assume that the only sizeable NP contributions enter  through $O_{7,8}^{(\prime)}$ which is the case for the models under consideration in this article. The operators $O_{7,8}^\prime$ are obtained by exchanging $L$ with $R$ and vice versa in the unprimed operators. The NLO decay width can thus be written as

\begin{equation}
\begin{split}
{\rm Br}\left[\bar{B} \to X_d \gamma\right] = {\cal N} \left|\frac{V_{td}^\star V_{tb}}{V_{cb}}\right|^2\left(P+N+P^\prime\right).
\end{split}
\end{equation}
Here $P$ contains the perturbative SM contributions and the NP contributions to $O_{7,8}$ while $P^\prime$ contains only the NP contributions to $O_{7,8}^\prime$. $N$ denotes the non-perturbative corrections and $\mathcal{N}\approx2.5\times 10^{-3}$ is a numerical prefactor (see Ref.~\cite{Hurth:2003dk} for details).

Before turning our attention to NP, we update the SM prediction for the CP averaged branching ratio $\left\langle \mathrm{Br}\left[ B\to X_d\gamma \right]_\mathrm{SM}\right\rangle$ of Refs.~\cite{Ali:1998rr,Hurth:2003dk} by using the improved determination of the CKM-element $V_{td}$ and the reduced non-perturbative uncertainties which are estimated to be at most $5\%$ (as for $B\to X_s\gamma$) \cite{Benzke:2010js,Hurth:2010tk}. The remaining leading uncertainty stems from renormalization scheme dependence of the ratio $m_c/m_b$ (approximately 15\%) which is supplemented by a $3.5\%$ scale ambiguity and a $6\%$ parametric uncertainty \cite{Hurth:2003dk}. In addition there is still a $10\%$ change in the branching ratio if one varies $V_{td}$ within its $95\%{\rm CL}$ region. Adding all these uncertainties in quadrature, we get:
\begin{equation}\label{NNLO_SM}
\left\langle \mathrm{Br}\left[ B\to X_d\gamma \right]^\mathrm{SM}_{E_{\gamma}>1.6 {\rm GeV}}\right\rangle = \;1.54^{+0.26}_{-0.31}\times10^{-5}.
\end{equation}
Comparing this with the experimental value of $\langle\mathrm{Br}\left[ B \to X_d \gamma \right]^\mathrm{exp}_{E_{\gamma}>1.6 {\rm GeV}}\rangle=(1.41\pm0.57) \times 10^{-5}$, we see that the SM prediction well within the experimental $1\sigma$ range. 

\begin{figure*}
\centering
\includegraphics[width=0.4\textwidth]{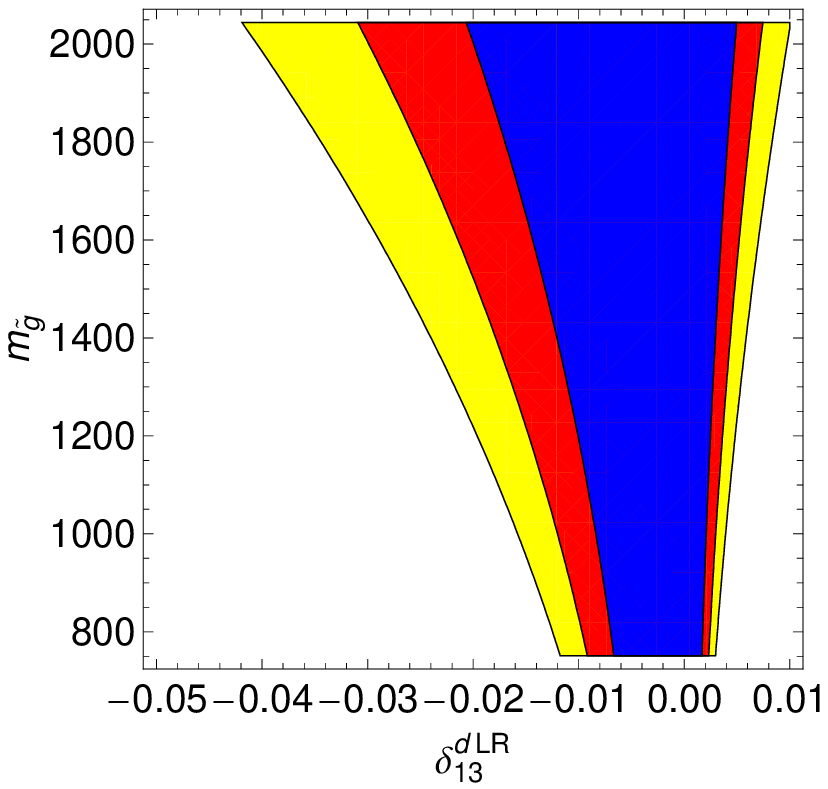}\hspace{0.5cm}
\includegraphics[width=0.4\textwidth]{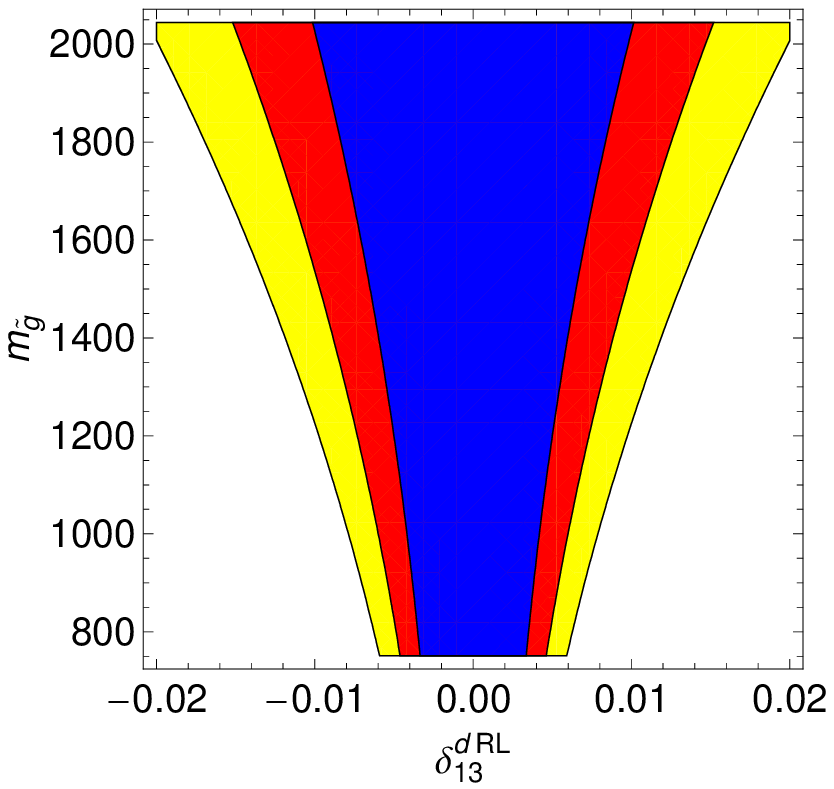}
\caption{Constraints on the mass insertion parameters $\delta_{13}^{d\;LR}$ (left plot) and $\delta_{13}^{d\;RL}$ (right plot) for $m_{\tilde g}=m_{\tilde q}=1\, \rm{TeV}$ and $\mu\tan\beta=30\,\mathrm{TeV}$ (yellow), $\mu\tan\beta=0$ (red), $\mu\tan\beta=-30\,\mathrm{TeV}$ (blue). Note that we only considered the leading gluino contributions (see Ref.~\cite{Crivellin:2009ar} for details). The constraints on $\delta_{13}^{d\;RL}$ are independent of its phase while the constraints on $\delta_{13}^{d\;LR}$ are given for $\mathrm{Arg}\left[\delta_{13}^{d\;RL}\right]=\mathrm{Arg}\left[V_{td}\right]$ and have to be scaled according to Fig.~\ref{C7} otherwise. In order to take into account the chirally enhanced corrections we used the effective Feynman rules of Ref.~\cite{Crivellin:2011jt}.
\label{deltaLR}}
\end{figure*}

\section{Constraints on new physics}

Since the SM contribution to $B\to X_d\gamma$ is not only loop- but also chirality-suppressed, this decay is, just as $B\to X_s\gamma$, very sensitive to new sources of flavor and chirality violation which occur in most NP models. In order to include NP into the calculation of the branching ratio we rely on the NLO formula of Ref.~\cite{Hurth:2003dk}.

The constraints in this section are obtained by demanding that branching ratio, including NP contributions, should lie within the $2\sigma$ range of the experimental values if not indicated otherwise. In order to give a conservative estimate we add the theory error and the experimental one linearly. Further we define:
\begin{equation}
C_{7,8}^{{\rm NP}}=C_{7,8}-C_{7,8}^{{\rm SM}},\qquad C_{7,8}^{\prime{\rm NP}}=C_{7,8}^\prime .
\end{equation}

\subsection{Model independent analysis}\label{sec modelindependent}

Firstly, we can constrain the Wilson coefficients $C_7^{{\rm NP}(\prime)}$ and $C_8^{{\rm NP}(\prime)}$ at the scale $M_W$.  
In the left plot of Fig.~\ref{C7} we show the $1\sigma$ and $2\sigma$ allowed region in the $Re[ C^{\mathrm{NP}}_{7}/C^{\mathrm{SM}}_{7}]$--$Im[C^{\mathrm{NP}}_{7}/C^{\mathrm{SM}}_{7}]$ plane for $C^{\mathrm{NP}}_7/C^{\mathrm{NP}}_8 = C^{\mathrm{SM}}_7 / C^{\mathrm{SM}}_8 $. Clearly the size of constructive contributions is very limited, but large destructive contributions are still possible. The primed operators always give a constructive contribution to the branching ratio and thus their possible size is rather limited (see right plot of Fig.~\ref{C7}). Note that $|C^{\mathrm{NP}}_{7,8} |$ can easily several times larger than $|C^{\mathrm{SM}}_{7,8}|$ in the case of destructing interference. The reason for this is that we normalize $C^{\mathrm{NP}}_{7,8}$ only to $C^{\mathrm{SM}}_{7,8}$, i.e. even for $C^{\mathrm{NP}}_{7,8} = - C^{\mathrm{SM}}_{7,8} $ the branching ratio is not zero because of the contributions from the SM four-quark operators. 

In models with minimal flavor violation (MFV) \cite{D'Ambrosio:2002ex}, the constraints on the Wilson coefficients $C_7^{{\rm NP}(\prime)}$ and $C_8^{{\rm NP}(\prime)}$ obtained in this section can be directly compared to the ones from $b\to s\gamma$ because the CKM elements are factored out in \eq{Heff}. Note that despite the recent improvements in $b\to d\gamma$ the constraints from $b\to s\gamma$ are still stronger if MFV is assumed.

\subsection{MSSM}

The generic MSSM possesses many new sources of flavor violation and constraining this flavor structure with FCNC processes has a long and fruitful tradition \cite{Bertolini:1990if,Hagelin:1992tc}. Concerning $B\to X_d\gamma$, we are especially sensitive to the chirality flipping elements $\delta^{d\,LR,RL}_{13}$ \cite{Ko:2002ee} (but also to $\delta^{d\,LL,RR}_{13}$ at moderate to large values of $\tan\beta$) and we get even more stringent constraints than from $B_d-\overline{B}_d$ mixing \cite{Becirevic:2001jj}. The results are depicted in Fig.~\ref{deltaLR}. 

\begin{figure*}
\centering
\includegraphics[width=0.4\textwidth]{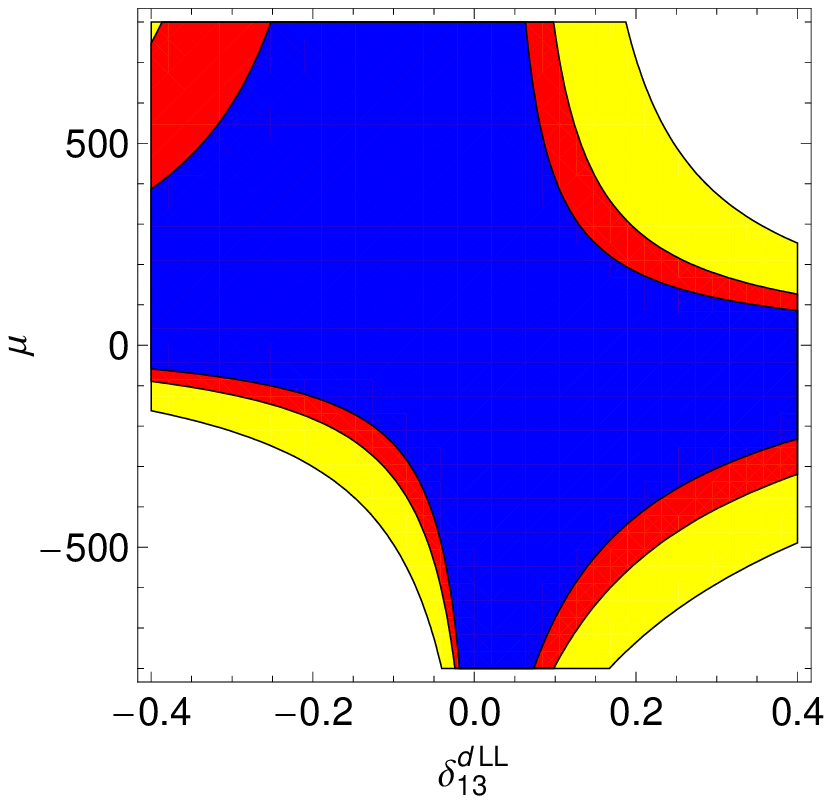}
\includegraphics[width=0.4\textwidth]{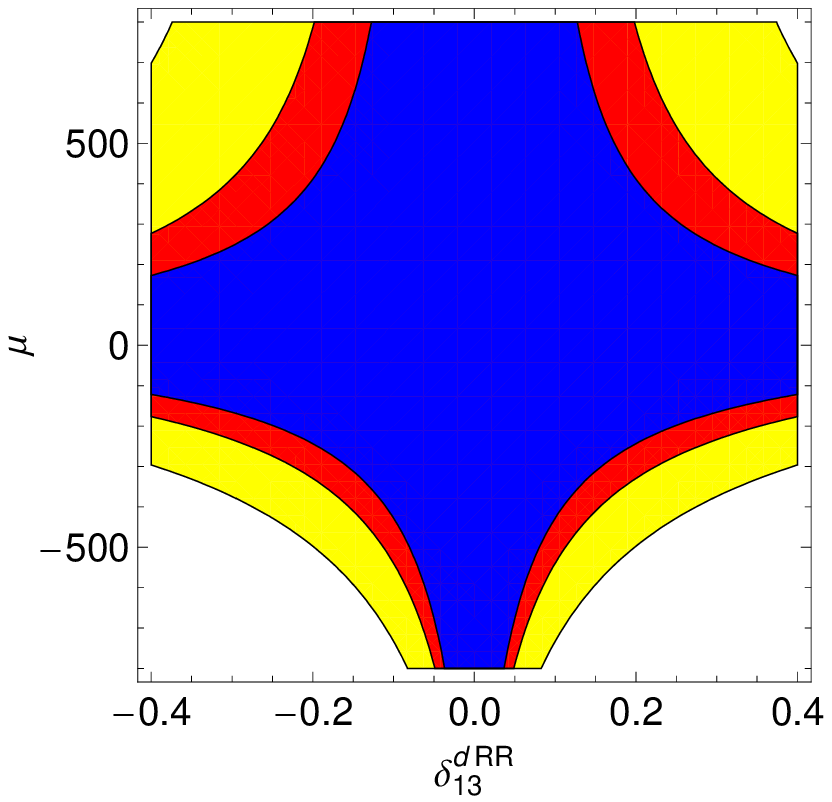}
\caption{Constraints on the mass insertion parameters $\delta_{13}^{d\;LR}$ (left plot) and $\delta_{13}^{d\;RL}$ (right plot) for $\tan\beta=50$, $m_{\tilde q}=1\, \rm{TeV}$ and $m_{\tilde g}=1.5\,\mathrm{TeV}$ (yellow), $m_{\tilde g}=1\,\mathrm{TeV}$ (red) $m_{\tilde g}=0.75\,\mathrm{TeV}$ (blue). The constraints on $\delta_{13}^{d\;LL}$ are independent of its phase while the constraints on $\delta_{13}^{d\;RR}$ are given for $\mathrm{Arg}\left[\delta_{13}^{d\;RR}\right]=\mathrm{Arg}\left[V_{td}\right]$ and have to be scaled according to Fig.~\ref{C7} if the phase is different.
\label{deltaLL}}
\end{figure*}

\subsection{2HDM of type III}

Despite the significant improvements in $B\to X_d\gamma$, the bounds from $B\to X_s \gamma$ are still tighter in scenarios with MFV \cite{D'Ambrosio:2002ex}. Thus the constraints on the charged Higgs mass of a two Higgs doublet models (2HDM) of type II are still more stringent from $B\to X_s \gamma$. However in a 2HDM of type III the non-holomorphic couplings of a $t$ and a $u$ quark to the Higgs can be constrained. This kind of models have been considered in Refs.~\cite{Buras:2010mh,Branco:2011iw,Gunion:1989we}, where however additional assumptions on the structure of the couplings has been imposed for the phenomenological studies.

Following the notation of Ref.~\cite{Crivellin:2010er} we denote the coupling coefficients of the charged Higgs vertex (for large $\tan\beta$) as 

%\begin{widetext}
\begin{equation}
\begin{array}{l}
{\Gamma _{u_f d_i }^{LR\;H^ \pm  } } =  \dfrac{1}{v}\sum\limits_{j = 1}^3 {V_{fj}^{CKM} \tan \left( \beta  \right)\left(  {   m_{d_i } \delta _{ji} -  \tilde \Sigma _{ji\;A^\prime\mu}^{ d\;LR} } \right)}  \\ \\
{\Gamma _{u_f d_i }^{RL\;H^ \pm  } }   =  \dfrac{1}{v}\sum\limits_{j= 1}^3 \Big(  {\tan \left( \beta  \right)}  \tilde \Sigma _{fj\;A^\prime \mu}^{u\;RL}  \\ \hspace{11ex} + \cot \left( \beta  \right)m_{u_f } \delta _{fj}  \Big) V_{ji}^{CKM}.
\end{array}  
\end{equation}
%\end{widetext}

\begin{figure}
\centering
\includegraphics[width=0.4\textwidth]{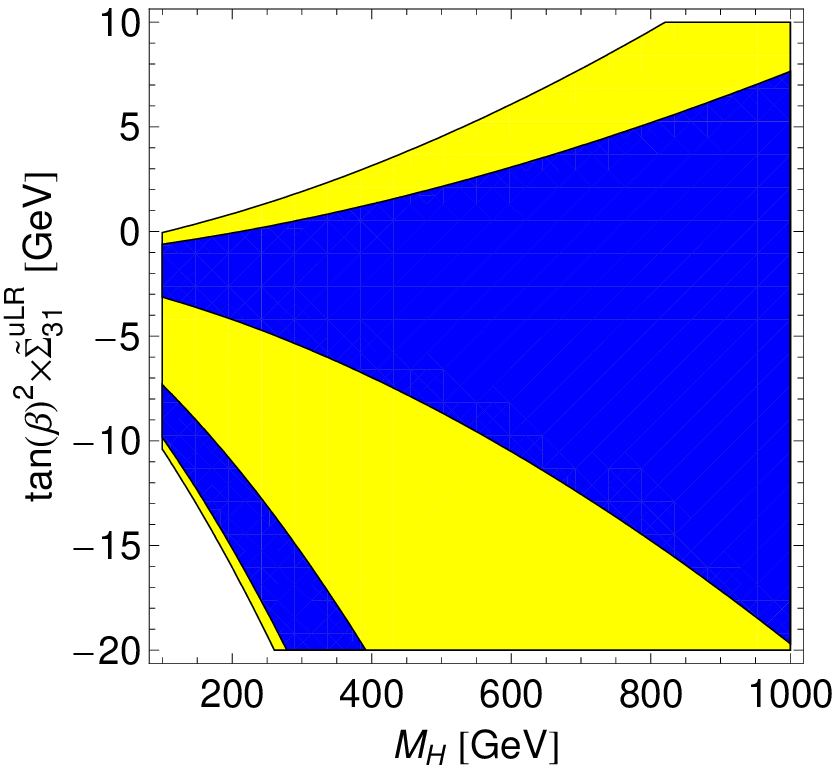}
\caption{Allowed regions in the ${\tan \left( \beta  \right)^2} \tilde \Sigma _{31\;A^\prime \mu}^{u\;RL}$--$M_{H^+}$ plane in the 2HDM III.
\label{Higgs}}
\end{figure}

In Fig.~\ref{Higgs} we show the constraints that we get on the product ${\tan \left( \beta  \right)^2} \tilde \Sigma _{31\;A^\prime \mu}^{u\;RL}$ which therefore, up to strongly suppressed terms, depend only on the charged Higgs mass. In principle we could also consider $\Sigma _{ji\;A^\prime\mu}^{ d\;LR}$, but the bounds from $B_d\to\mu^+\mu^-$ are more stringent.

\subsection{Right-handed charged currents}

It is well known that $B\to X_s \gamma$ puts stringent constraints on models with right-handed charged currents \cite{Cho:1993zb,Grzadkowski:2008mf}. We can thus also constrain the elements of the right-handed mixing matrix through $B\to X_d \gamma$ \footnote{An effective right-handed $W$-coupling can also be induced in the MSSM \cite{Crivellin:2009sd} which then affects the determination of $V_{ub}$ and $V_{cb}$. However, in this case $B\to X_d\gamma$ is not dangerous because one cannot generate a sizable $V_{td}^R$ coupling.}. We define the effective $W$-quark-quark vertex as
\begin{equation}
	i\Gamma^{W\,\mu}_{t,d}=-i\frac{g_2}{\sqrt{2}}\gamma^\mu\left(V_{td}P_L+V_{td}^R P_R\right) \;.
\end{equation}
If $V_{td}^R\neq0$ contributions to $C^\prime_{7,8}$ are induced which necessarily enhance the branching ratio. Using the formulas of Ref. \cite{Grzadkowski:2008mf} and assuming the SM value for $V_{td}$, we get the following limit on $V_{td}^R$
\begin{equation}
| V_{td}^R | \leq 1.5\times10^{-4} \;.
\end{equation}
Note that this constraint is approximately $3.5$ times stronger than what is found for the best-fit solution of the right-handed CKM matrix in Ref.~\cite{Buras:2010pz}.

\section{Conclusions and outlook}

In this letter we studied the constraints on NP from the inclusive radiative decay $B\to X_d\gamma$. Including the improved determination of $V_{td}$ and the reduced hadronic uncertainties \cite{Hurth:2010tk,Benzke:2010js} in the CP-averaged branching ratio, the new NLO SM prediction is given by $\langle\mathrm{Br}\left[ B\to X_d\gamma \right]^\mathrm{SM}_{E_{\gamma}>1.6 {\rm GeV}}\rangle = \;1.54^{+0.26}_{-0.31}\times10^{-5}$. If we extrapolate the experimental value from the BABAR collaboration \cite{BABAR:2010ps} to a photon energy cut of 1.6 GeV, we get $\langle\mathrm{Br}\left[ B \to X_d \gamma \right]^\mathrm{exp}_{E_{\gamma}>1.6 {\rm GeV}}\rangle=(1.41\pm0.57) \times 10^{-5}$.  

We found constraints on the parameters $\delta^{d\,LR,RL}_{13}$ of the MSSM squark mass matrices which are more stringent than the ones obtained from \bbm. Also, an effective right-handed $W$ coupling to the top and down quarks is severely constrained: $| V_{td}^R | \leq 1.5\times10^{-4}$. 
This for example strongly disfavors the proposed best-fit solution to the right-handed CKM matrix of Ref.~\cite{Buras:2010pz}.

The significance of $B\to X_d\gamma$ can even be improved by a NNLO computation of the SM prediction which is in progress. In addition, an analysis of the existing BELLE data would be welcome in order to reduce the error of the measurement. $B\to X_d\gamma$ is also very interesting for future super-B factories which will be able to measure this decay very precisely.

%%%%%%%%%%%%%%%%%%%%%%%%%%%%%%
\vspace{7mm}
{\it Acknowledgments.}--- 
We are grateful to C. Greub, L. Hofer and M. Procura for proofreading the article and M. Misiak for useful comments on the manuscript. A.C. thanks C. Greub, M. Misiak and U. Nierste  for useful discussions and W. Altmannshofer and M. Misiak for pointing out that the HFAG value, used in the first version of this article, was not extrapolated to a photon energy cut of 1.6 GeV. This work is supported by the Swiss National Foundation. The Albert Einstein Center for Fundamental Physics
is supported by the ``Innovations- und Kooperationsprojekt C-13 of the
Schweizerische Universit\"atskonferenz SUK/CRUS''.

\bibliography{b-d-gamma}% Produces the bibliography via BibTeX.

\begin{thebibliography}{29}
\expandafter\ifx\csname natexlab\endcsname\relax\def\natexlab#1{#1}\fi
\expandafter\ifx\csname bibnamefont\endcsname\relax
  \def\bibnamefont#1{#1}\fi
\expandafter\ifx\csname bibfnamefont\endcsname\relax
  \def\bibfnamefont#1{#1}\fi
\expandafter\ifx\csname citenamefont\endcsname\relax
  \def\citenamefont#1{#1}\fi
\expandafter\ifx\csname url\endcsname\relax
  \def\url#1{\texttt{#1}}\fi
\expandafter\ifx\csname urlprefix\endcsname\relax\def\urlprefix{URL }\fi
\providecommand{\bibinfo}[2]{#2}
\providecommand{\eprint}[2][]{\url{#2}}

\bibitem[{\citenamefont{Nakamura et~al.}(2010)}]{Nakamura:2010zzi}
\bibinfo{author}{\bibfnamefont{K.}~\bibnamefont{Nakamura}} \bibnamefont{et~al.}
  (\bibinfo{collaboration}{Particle Data Group}), \bibinfo{journal}{J. Phys.}
  \textbf{\bibinfo{volume}{G37}}, \bibinfo{pages}{075021}
  (\bibinfo{year}{2010}).

\bibitem[{\citenamefont{Misiak et~al.}(2007)}]{Misiak:2006zs}
\bibinfo{author}{\bibfnamefont{M.}~\bibnamefont{Misiak}} \bibnamefont{et~al.},
  \bibinfo{journal}{Phys. Rev. Lett.} \textbf{\bibinfo{volume}{98}},
  \bibinfo{pages}{022002} (\bibinfo{year}{2007}), \eprint{hep-ph/0609232}.

\bibitem[{\citenamefont{Misiak and Steinhauser}(2007)}]{Misiak:2006ab}
\bibinfo{author}{\bibfnamefont{M.}~\bibnamefont{Misiak}} \bibnamefont{and}
  \bibinfo{author}{\bibfnamefont{M.}~\bibnamefont{Steinhauser}},
  \bibinfo{journal}{Nucl. Phys.} \textbf{\bibinfo{volume}{B764}},
  \bibinfo{pages}{62} (\bibinfo{year}{2007}), \eprint{hep-ph/0609241}.

\bibitem[{\citenamefont{del Amo~Sanchez et~al.}(2010)}]{BABAR:2010ps}
\bibinfo{author}{\bibfnamefont{P.}~\bibnamefont{del Amo~Sanchez}}
  \bibnamefont{et~al.} (\bibinfo{collaboration}{BABAR}),
  \bibinfo{journal}{Phys. Rev.} \textbf{\bibinfo{volume}{D82}},
  \bibinfo{pages}{051101} (\bibinfo{year}{2010}), \eprint{1005.4087}.

\bibitem[{\citenamefont{Wang}(2011)}]{Wang:2011sn}
\bibinfo{author}{\bibfnamefont{W.}~\bibnamefont{Wang}} (\bibinfo{year}{2011}),
  \eprint{1102.1925}.

\bibitem[{\citenamefont{O'Leary et~al.}(2010)}]{O'Leary:2010af}
\bibinfo{author}{\bibfnamefont{B.}~\bibnamefont{O'Leary}} \bibnamefont{et~al.}
  (\bibinfo{collaboration}{SuperB}) (\bibinfo{year}{2010}), \eprint{1008.1541}.

\bibitem[{\citenamefont{Abe et~al.}(2010)}]{Abe:2010sj}
\bibinfo{author}{\bibfnamefont{T.}~\bibnamefont{Abe}} \bibnamefont{et~al.}
  (\bibinfo{collaboration}{Belle II}) (\bibinfo{year}{2010}),
  \eprint{1011.0352}.

\bibitem[{\citenamefont{Ali and Greub}(1992)}]{Ali:1992qs}
\bibinfo{author}{\bibfnamefont{A.}~\bibnamefont{Ali}} \bibnamefont{and}
  \bibinfo{author}{\bibfnamefont{C.}~\bibnamefont{Greub}},
  \bibinfo{journal}{Phys. Lett.} \textbf{\bibinfo{volume}{B287}},
  \bibinfo{pages}{191} (\bibinfo{year}{1992}).

\bibitem[{\citenamefont{Ali et~al.}(1998)\citenamefont{Ali, Asatrian, and
  Greub}}]{Ali:1998rr}
\bibinfo{author}{\bibfnamefont{A.}~\bibnamefont{Ali}},
  \bibinfo{author}{\bibfnamefont{H.}~\bibnamefont{Asatrian}}, \bibnamefont{and}
  \bibinfo{author}{\bibfnamefont{C.}~\bibnamefont{Greub}},
  \bibinfo{journal}{Phys. Lett.} \textbf{\bibinfo{volume}{B429}},
  \bibinfo{pages}{87} (\bibinfo{year}{1998}), \eprint{hep-ph/9803314}.

\bibitem[{\citenamefont{Hurth and Nakao}(2010)}]{Hurth:2010tk}
\bibinfo{author}{\bibfnamefont{T.}~\bibnamefont{Hurth}} \bibnamefont{and}
  \bibinfo{author}{\bibfnamefont{M.}~\bibnamefont{Nakao}},
  \bibinfo{journal}{Ann. Rev. Nucl. Part. Sci.} \textbf{\bibinfo{volume}{60}},
  \bibinfo{pages}{645} (\bibinfo{year}{2010}), \eprint{1005.1224}.

\bibitem[{\citenamefont{Benzke et~al.}(2010)\citenamefont{Benzke, Lee, Neubert,
  and Paz}}]{Benzke:2010js}
\bibinfo{author}{\bibfnamefont{M.}~\bibnamefont{Benzke}},
  \bibinfo{author}{\bibfnamefont{S.~J.} \bibnamefont{Lee}},
  \bibinfo{author}{\bibfnamefont{M.}~\bibnamefont{Neubert}}, \bibnamefont{and}
  \bibinfo{author}{\bibfnamefont{G.}~\bibnamefont{Paz}},
  \bibinfo{journal}{JHEP} \textbf{\bibinfo{volume}{08}}, \bibinfo{pages}{099}
  (\bibinfo{year}{2010}), \eprint{1003.5012}.

\bibitem[{\citenamefont{Charles et~al.}(2005)}]{Charles:2004jd}
\bibinfo{author}{\bibfnamefont{J.}~\bibnamefont{Charles}} \bibnamefont{et~al.}
  (\bibinfo{collaboration}{CKMfitter Group}), \bibinfo{journal}{Eur. Phys. J.}
  \textbf{\bibinfo{volume}{C41}}, \bibinfo{pages}{1} (\bibinfo{year}{2005}),
  \eprint{hep-ph/0406184}.

\bibitem[{\citenamefont{Ciuchini et~al.}(2001)}]{Ciuchini:2000de}
\bibinfo{author}{\bibfnamefont{M.}~\bibnamefont{Ciuchini}}
  \bibnamefont{et~al.}, \bibinfo{journal}{JHEP} \textbf{\bibinfo{volume}{07}},
  \bibinfo{pages}{013} (\bibinfo{year}{2001}), \eprint{hep-ph/0012308}.

\bibitem[{\citenamefont{Hurth et~al.}(2005)\citenamefont{Hurth, Lunghi, and
  Porod}}]{Hurth:2003dk}
\bibinfo{author}{\bibfnamefont{T.}~\bibnamefont{Hurth}},
  \bibinfo{author}{\bibfnamefont{E.}~\bibnamefont{Lunghi}}, \bibnamefont{and}
  \bibinfo{author}{\bibfnamefont{W.}~\bibnamefont{Porod}},
  \bibinfo{journal}{Nucl. Phys.} \textbf{\bibinfo{volume}{B704}},
  \bibinfo{pages}{56} (\bibinfo{year}{2005}), \eprint{hep-ph/0312260}.

\bibitem[{\citenamefont{Crivellin and Nierste}(2009)}]{Crivellin:2009ar}
\bibinfo{author}{\bibfnamefont{A.}~\bibnamefont{Crivellin}} \bibnamefont{and}
  \bibinfo{author}{\bibfnamefont{U.}~\bibnamefont{Nierste}}
  (\bibinfo{year}{2009}), \eprint{0908.4404}.

\bibitem[{\citenamefont{Crivellin et~al.}(2011)\citenamefont{Crivellin, Hofer,
  and Rosiek}}]{Crivellin:2011jt}
\bibinfo{author}{\bibfnamefont{A.}~\bibnamefont{Crivellin}},
  \bibinfo{author}{\bibfnamefont{L.}~\bibnamefont{Hofer}}, \bibnamefont{and}
  \bibinfo{author}{\bibfnamefont{J.}~\bibnamefont{Rosiek}}
  (\bibinfo{year}{2011}), \eprint{1103.4272}.

\bibitem[{\citenamefont{D'Ambrosio et~al.}(2002)\citenamefont{D'Ambrosio,
  Giudice, Isidori, and Strumia}}]{D'Ambrosio:2002ex}
\bibinfo{author}{\bibfnamefont{G.}~\bibnamefont{D'Ambrosio}},
  \bibinfo{author}{\bibfnamefont{G.~F.} \bibnamefont{Giudice}},
  \bibinfo{author}{\bibfnamefont{G.}~\bibnamefont{Isidori}}, \bibnamefont{and}
  \bibinfo{author}{\bibfnamefont{A.}~\bibnamefont{Strumia}},
  \bibinfo{journal}{Nucl. Phys.} \textbf{\bibinfo{volume}{B645}},
  \bibinfo{pages}{155} (\bibinfo{year}{2002}), \eprint{hep-ph/0207036}.

\bibitem[{\citenamefont{Bertolini et~al.}(1991)\citenamefont{Bertolini,
  Borzumati, Masiero, and Ridolfi}}]{Bertolini:1990if}
\bibinfo{author}{\bibfnamefont{S.}~\bibnamefont{Bertolini}},
  \bibinfo{author}{\bibfnamefont{F.}~\bibnamefont{Borzumati}},
  \bibinfo{author}{\bibfnamefont{A.}~\bibnamefont{Masiero}}, \bibnamefont{and}
  \bibinfo{author}{\bibfnamefont{G.}~\bibnamefont{Ridolfi}},
  \bibinfo{journal}{Nucl. Phys.} \textbf{\bibinfo{volume}{B353}},
  \bibinfo{pages}{591} (\bibinfo{year}{1991}).

\bibitem[{\citenamefont{Hagelin et~al.}(1994)\citenamefont{Hagelin, Kelley, and
  Tanaka}}]{Hagelin:1992tc}
\bibinfo{author}{\bibfnamefont{J.~S.} \bibnamefont{Hagelin}},
  \bibinfo{author}{\bibfnamefont{S.}~\bibnamefont{Kelley}}, \bibnamefont{and}
  \bibinfo{author}{\bibfnamefont{T.}~\bibnamefont{Tanaka}},
  \bibinfo{journal}{Nucl. Phys.} \textbf{\bibinfo{volume}{B415}},
  \bibinfo{pages}{293} (\bibinfo{year}{1994}).

\bibitem[{\citenamefont{Ko et~al.}(2002)\citenamefont{Ko, Park, and
  Kramer}}]{Ko:2002ee}
\bibinfo{author}{\bibfnamefont{P.}~\bibnamefont{Ko}},
  \bibinfo{author}{\bibfnamefont{J.-h.} \bibnamefont{Park}}, \bibnamefont{and}
  \bibinfo{author}{\bibfnamefont{G.}~\bibnamefont{Kramer}},
  \bibinfo{journal}{Eur. Phys. J.} \textbf{\bibinfo{volume}{C25}},
  \bibinfo{pages}{615} (\bibinfo{year}{2002}), \eprint{hep-ph/0206297}.

\bibitem[{\citenamefont{Becirevic et~al.}(2002)}]{Becirevic:2001jj}
\bibinfo{author}{\bibfnamefont{D.}~\bibnamefont{Becirevic}}
  \bibnamefont{et~al.}, \bibinfo{journal}{Nucl. Phys.}
  \textbf{\bibinfo{volume}{B634}}, \bibinfo{pages}{105} (\bibinfo{year}{2002}),
  \eprint{hep-ph/0112303}.

\bibitem[{\citenamefont{Buras et~al.}(2010)\citenamefont{Buras, Carlucci, Gori,
  and Isidori}}]{Buras:2010mh}
\bibinfo{author}{\bibfnamefont{A.~J.} \bibnamefont{Buras}},
  \bibinfo{author}{\bibfnamefont{M.~V.} \bibnamefont{Carlucci}},
  \bibinfo{author}{\bibfnamefont{S.}~\bibnamefont{Gori}}, \bibnamefont{and}
  \bibinfo{author}{\bibfnamefont{G.}~\bibnamefont{Isidori}},
  \bibinfo{journal}{JHEP} \textbf{\bibinfo{volume}{1010}}, \bibinfo{pages}{009}
  (\bibinfo{year}{2010}), \eprint{1005.5310}.

\bibitem[{\citenamefont{Branco et~al.}(2011)\citenamefont{Branco, Ferreira,
  Lavoura, Rebelo, Sher et~al.}}]{Branco:2011iw}
\bibinfo{author}{\bibfnamefont{G.}~\bibnamefont{Branco}},
  \bibinfo{author}{\bibfnamefont{P.}~\bibnamefont{Ferreira}},
  \bibinfo{author}{\bibfnamefont{L.}~\bibnamefont{Lavoura}},
  \bibinfo{author}{\bibfnamefont{M.}~\bibnamefont{Rebelo}},
  \bibinfo{author}{\bibfnamefont{M.}~\bibnamefont{Sher}}, \bibnamefont{et~al.}
  (\bibinfo{year}{2011}), \bibinfo{note}{* Temporary entry *},
  \eprint{1106.0034}.

\bibitem[{\citenamefont{Gunion et~al.}(2000)\citenamefont{Gunion, Haber, Kane,
  and Dawson}}]{Gunion:1989we}
\bibinfo{author}{\bibfnamefont{J.~F.} \bibnamefont{Gunion}},
  \bibinfo{author}{\bibfnamefont{H.~E.} \bibnamefont{Haber}},
  \bibinfo{author}{\bibfnamefont{G.~L.} \bibnamefont{Kane}}, \bibnamefont{and}
  \bibinfo{author}{\bibfnamefont{S.}~\bibnamefont{Dawson}},
  \bibinfo{journal}{Front.Phys.} \textbf{\bibinfo{volume}{80}},
  \bibinfo{pages}{1} (\bibinfo{year}{2000}).

\bibitem[{\citenamefont{Crivellin}(2011)}]{Crivellin:2010er}
\bibinfo{author}{\bibfnamefont{A.}~\bibnamefont{Crivellin}},
  \bibinfo{journal}{Phys. Rev.} \textbf{\bibinfo{volume}{D83}},
  \bibinfo{pages}{056001} (\bibinfo{year}{2011}), \eprint{1012.4840}.

\bibitem[{\citenamefont{Cho and Misiak}(1994)}]{Cho:1993zb}
\bibinfo{author}{\bibfnamefont{P.~L.} \bibnamefont{Cho}} \bibnamefont{and}
  \bibinfo{author}{\bibfnamefont{M.}~\bibnamefont{Misiak}},
  \bibinfo{journal}{Phys. Rev.} \textbf{\bibinfo{volume}{D49}},
  \bibinfo{pages}{5894} (\bibinfo{year}{1994}), \eprint{hep-ph/9310332}.

\bibitem[{\citenamefont{Grzadkowski and Misiak}(2008)}]{Grzadkowski:2008mf}
\bibinfo{author}{\bibfnamefont{B.}~\bibnamefont{Grzadkowski}} \bibnamefont{and}
  \bibinfo{author}{\bibfnamefont{M.}~\bibnamefont{Misiak}},
  \bibinfo{journal}{Phys. rev.} \textbf{\bibinfo{volume}{D78}},
  \bibinfo{pages}{077501} (\bibinfo{year}{2008}), \eprint{0802.1413}.

\bibitem[{\citenamefont{Crivellin}(2010)}]{Crivellin:2009sd}
\bibinfo{author}{\bibfnamefont{A.}~\bibnamefont{Crivellin}},
  \bibinfo{journal}{Phys. Rev.} \textbf{\bibinfo{volume}{D81}},
  \bibinfo{pages}{031301} (\bibinfo{year}{2010}), \eprint{0907.2461}.

\bibitem[{\citenamefont{Buras et~al.}(2011)\citenamefont{Buras, Gemmler, and
  Isidori}}]{Buras:2010pz}
\bibinfo{author}{\bibfnamefont{A.~J.} \bibnamefont{Buras}},
  \bibinfo{author}{\bibfnamefont{K.}~\bibnamefont{Gemmler}}, \bibnamefont{and}
  \bibinfo{author}{\bibfnamefont{G.}~\bibnamefont{Isidori}},
  \bibinfo{journal}{Nucl. Phys.} \textbf{\bibinfo{volume}{B843}},
  \bibinfo{pages}{107} (\bibinfo{year}{2011}), \eprint{1007.1993}.

\end{thebibliography}

\end{document}